\documentclass[12pt]{iopart}
\usepackage{graphicx}

\begin{document}

\title{Pair production from nuclear collisions and cosmic ray transport}

\author{John W. Norbury}

\address{Department of Physics and Astronomy, University of Southern Mississippi, Hattiesburg, Mississippi, 39402, USA}
\ead{john.norbury@usm.edu}
\begin{abstract}
Modern cosmic ray transport codes, that are capable of use for a variety of applications, need to include all significant atomic, nuclear and particle reactions at a variety of energies. Lepton pair production from nucleus-nucleus collisions  has not been included in transport codes to date. Using the methods of Baur, Bertulani and Baron, the present paper provides estimates of electron-positron pair production cross sections for nuclei and energies relevant to cosmic ray transport. It is shown that the cross sections are   large compared to other typical processes such as single neutron removal due to strong or electromagnetic interactions. Therefore lepton pair production  may need to  be included in some  transport code applications involving  MeV electrons.

\end{abstract}



\section{Introduction}

Cosmic radiation transport codes \cite{wilson} find use in many diverse areas of physics, astrophysics and space applications. By transporting protons and nuclei through the interstellar medium and comparing the results to the cosmic ray spectrum observed in the solar system, one can deduce properties of interstellar space and also learn about the primary cosmic ray sources. In turn by propagating particles through the Earth atmosphere one can relate ground based observations to the solar system cosmic ray spectrum. This is especially true of extremely high energy cosmic rays where the flux is so low that only ground based observations are feasible. 
(Actually, a downward looking telescope viewing the fluorescence radiation of a cosmic ray shower trail from the vantage of the space station or from a free flyer gives a much better way to detect these highest energy cosmic rays.)
Cosmic transport codes also find many applications in aerospace. It is important to be able to predict radiation environments inside satellites so that sensitive electronics can be protected. The same is true for robot missions to planets such as Mars, Saturn and Jupiter. Human space flight in low Earth orbit and future Lunar and Martian missions also require an excellent knowledge of radiation environments inside the spacecraft. Again these environments are predicted by transporting cosmic ray particles through spacecraft walls.

The incident cosmic ray spectrum \cite{simpson}  consists of protons, leptons and heavier nuclei. The peak of the spectrum lies in the MeV - GeV region. Typical targets through which particles are transported contain protons and heavier nuclei. Therefore a cosmic ray transport code must include   all   the atomic, nuclear and particle physics pertaining to an incident nucleus impinging upon any target nucleus over a wide range of energies. A huge knowledge database is therefore required which must include many   reactions. Of course, depending on the applications, some reactions may not be as important as others. However, the best transport codes available today, such as HZETRN \cite{hzetrn}, GEANT \cite{geant} and FLUKA  \cite{fluka} aim at including all possible physics so that they can be used for any application ranging from design of accelerator experiments to radiation therapy. Including as large a knowledge database as possible is also important for space applications because the code may be used for such wide ranging topics as calculating the radiation environment inside a spacesuit  to the environment on the moon Callisto.

\section{Electron-positron production}

A reaction that has not been included in transport codes up to now is lepton pair production in nucleus-nucleus collisions, denoted as 
\begin{eqnarray}
A_P+A_T \rightarrow A_P+A_T +e^+ +e^-   \label{reaction}
\end{eqnarray}
for the case where the lepton pair consists of electrons and positrons. Here $A_P$  and $A_T$ are the mass numbers of the projectile and target nucleus respectively. Diagrams  illustrating this process are shown in figure 1.  \\

\includegraphics[width=5in]{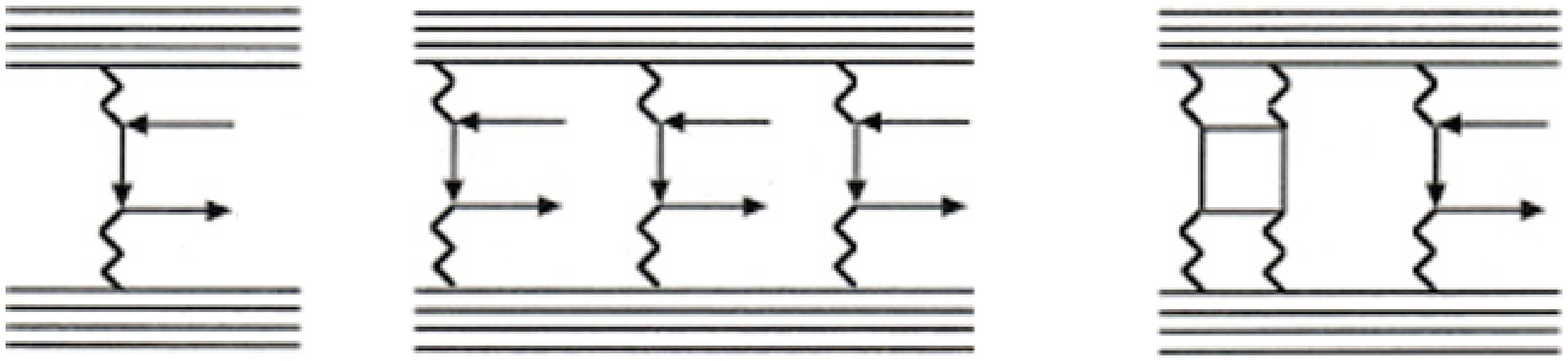}  

\noindent {\bf Figure 1.}   Some Feynman diagrams for electron-positron pair production via two-photon interactions generated through  peripheral nuclear collisions. The projectile and target nuclei are shown as  horizontal straight lines.\\

This reaction has been studied in detail during the last two decades  \cite{bertulani, baur2} and has lead to the realizations that the cross sections are very large \cite{bertulani}. The spectrum of the resulting electrons has also been calculated and typical electron energies are in the MeV region  \cite{baron}. Applications to spaceflight  immediately come to mind. It is known \cite{nrc}  that a 
10 MeV proton can successfully penetrate  a spacesuit (typically of density  ${\rm 0.5 \; g/cm^2}$  of equivalent Al)
 and a 
30 MeV proton can penetrate the  mid-deck of the space shuttle. In comparison electrons are easily shielded and  
electrons in the MeV range don't penetrate spacecraft such as the shuttle. However it is also known that a 
0.5 MeV electron can penetrate a spacesuit and therefore in calculating radiation environments for extra-vehicular activity (EVA), or other applications involving MeV electrons, one may need to  consider the reaction of equation (\ref{reaction}). 

The aim of the present paper is to give some typical cross section estimates for pair production processes that are relevant to the nuclei and energies found in the cosmic ray spectrum. These are shown in Table 1. 
For the sake of comparison, the cross sections for single neutron removal via strong and electromagnetic interactions are also shown. $\sigma_{Strong}$, $\sigma_{EM}$  and $\sigma_{pair}$  are  calculated according to the methods of References 
 \cite{bcv}, \cite{ norbury} and \cite{ baron, baur} respectively.  The equation for lepton pair production is \cite{baron, baur}
\begin{eqnarray}
\sigma 
&=&  \frac{28}{27 \pi} \left  (\frac{Z_P  Z_T \alpha^2 }{m_l} \right )^2\log^3 (\frac{\gamma }{2})    
\end{eqnarray}
where $Z_P$ and $Z_T$ are the projectile and target charges, $\alpha$ is the fine structure constant, $m_l$  is the lepton mass and $\gamma$ is the relativistic gamma factor of the projectile.  The various factors can be understood from the first order diagram as follows. Each photon-nucleus  vertex contains $Z \sqrt{\alpha}$, where $\alpha$ is the fine structure constant,  and each electron-photon vertex contains $\sqrt{\alpha}$.

From Table 1 it can be seen that in comparison to the neutron removal cross sections which are typically in the  millibarn  range, the pair production cross sections  are much larger, typically in the barn  region. In fact the pair cross section for Au + Fe  at 50 GeV is about 200 barn. The reason is that  pair production cross sections scale as $Z_P^2 Z_T^2$   \cite{bertulani}. The cross sections for production of heavier leptons, such as muons, are much smaller, due to the factor $1/m_l^2$.

\section{Conclusions}
Lepton pair production cross sections have been calculated in the past and have been shown to be large \cite{bertulani, baur2, baron, baur}. However these calculations have typically been performed at very high energies and for very heavy nuclei.
The aim of the present work has been to present these  cross sections  for the nuclei and energies typically found in the cosmic ray spectrum and to compare them to other typical reaction cross sections.
Compared to single nucleon removal, it is seen that electron-positron pair production cross sections from nuclear collisions are  large for the nuclei and energies typically found in the cosmic ray spectrum. Therefore it  might be necessary to include  this reaction in cosmic ray transport codes  involving MeV electrons.  This will depend on the particular application, the particles of interest and the incident spectrum that one is considering.\\

\noindent Acknowledgements: 
JWN   was supported by
NASA Grant NNL05AA05G.

\newpage

\noindent {\bf Table 1.}  Pair production cross sections $\sigma_{pair}$  from nuclear collisions. $\sigma_{Strong}$ and $\sigma_{EM}$  denote the strong and electromagnetic  cross sections for single neutron removal. $T$ is the kinetic energy of the projectile.

\noindent
\hrulefill
\begin{tabbing}
xxxxxxxxxxxx\=xxxxxxxxxxxx\=xxxxxxxxxxxxxx\=xxxxxxxxxxxx\=xxxxxxxxxxxx\=xxxxxxxxxxxxx\kill
Projectile\> Target \>T  \>$\sigma_{\rm Strong}$  \>$\sigma_{\rm EM}$\>$\sigma_{\rm pair}$   \\
\>  \> (AGeV) \> (mb) \> (mb) \> (mb) 
\end{tabbing}
\hrulefill
\begin{tabbing}
xxxxxxxxxxxx\=xxxxxxxxxxxx\=xxxxxxxxxxxxxx\=xxxxxxxxxxxx\=xxxxxxxxxxxx\=xxxxxxxxxxxxx\kill
$^{12}$C      \>$^{12}$C    \>3     \>64      \>0.6  \>0.7      \\
              \>            \>5     \>64      \> 0.7    \> 2.8     \\
              \>            \>10     \>64      \>0.9     \> 10              \\
              \>            \>50     \>64      \>1.5     \>  65             \\\\
              \>$^{27}$Al    \>3     \>77      \>2.4     \>3               \\
              \>             \>5     \>77      \>3      \>13              \\
              \>             \>10     \>77      \>4        \> 46           \\
              \>             \>50     \>77      \>6       \> 305            \\\\
              \>$^{56}$Fe    \>3     \>92        \>8.5        \>14            \\
              \>             \>5     \>92        \>11        \> 52           \\
              \>             \>10     \>92        \>15        \>186             \\
              \>             \>50     \>92        \>25        \>1220        \\  \\ 
$^{28}$Si      \>$^{12}$C    \>3     \>73      \>1.1       \> 4            \\
              \>            \>5     \>73      \> 1.4        \> 15          \\
              \>            \>10     \>73        \> 1.8       \> 54           \\
              \>            \>50     \>73        \> 3        \>354           \\\\
              \>$^{27}$Al    \>3     \>86        \>5         \> 19          \\
              \>             \>5     \>86        \>6       \> 71           \\
              \>             \>10     \>86        \>8       \> 253            \\
              \>             \>50     \>86        \>14        \>1661             \\\\
              \>$^{56}$Fe    \>3     \>100        \>17      \>75               \\
              \>             \>5     \>100        \>22         \>282           \\
              \>             \>10     \>100        \>30       \>  1011           \\
              \>             \>50     \>100       \> 52       \>6643    \\\\
$^{56}$Fe      \>$^{12}$C    \>3     \>89      \>  7       \> 14         \\
              \>            \>5     \>89        \> 8          \> 52        \\
              \>            \>10     \>89        \> 11        \> 186           \\
              \>            \>50     \>89        \> 17         \>1220          \\\\
              \>$^{27}$Al    \>3     \>102        \> 27      \> 65             \\
              \>             \>5     \> 102       \> 36        \>243           \\
              \>             \>10     \>102        \> 47        \> 872           \\
              \>             \>50     \> 102       \> 77        \> 5728           \\\\
              \>$^{56}$Fe    \>3     \>116          \> 104       \>259             \\
              \>             \>5     \>116          \> 132        \>973          \\
              \>             \>10     \>116          \> 178        \> 3487          \\
              \>             \>50     \>116          \> 298         \>22910           \\\\
$^{197}$Au      \>$^{12}$C    \>3     \>128      \>  46       \> 127           \\
              \>            \>5     \>  128      \>  56         \>478          \\
              \>            \>10     \> 128       \>73         \>1714            \\
              \>            \>50     \> 128       \> 118       \>11264            \\\\
              \>$^{27}$Al    \>3     \>141        \> 201         \>598           \\
              \>             \>5     \>141        \> 250          \> 2245         \\
              \>             \>10     \> 141       \> 330        \> 8048          \\
              \>             \>50     \>141        \> 541        \>  52879          \\\\
              \>$^{56}$Fe    \>3     \>156          \> 749       \> 2391           \\
              \>             \>5     \>156          \> 946        \>8980           \\
              \>             \>10     \>156          \> 1263      \> 32193             \\
              \>             \>50     \>156          \> 2108       \>211516             
\end{tabbing}
\hrulefill

\end{document}